\documentclass[utf8]{frontiersinFPHY_FAMS}

\setcitestyle{square}
\usepackage{url,hyperref,lineno,microtype,subcaption,siunitx}
\usepackage[onehalfspacing]{setspace}

\def\keyFont{\fontsize{8}{11}\helveticabold }
\def\firstAuthorLast{Schütze {et~al.}} 
\def\Authors{Paul Schütze\,$^{1,*}$, Aenne Abel\,$^{1,2}$, Florian Burkart\,$^{1}$, L. Malinda S. de Silva\,$^{1}$, Hannes Dinter\,$^{1}$, Kevin Dojan\,$^{3}$, Adrian Herkert\,$^{1}$, Sonja Jaster-Merz\,$^{1}$, Max Joseph Kellermeier\,$^{1}$, Willi Kuropka\,$^{1}$, Frank Mayet\,$^{1}$, Sara Ruiz Daza\,$^{1,3}$, Simon Spannagel\,$^{1}$,  Thomas Vinatier\,$^{1}$, H\aa kan Wennlöf\,$^{1}$}
\def\Address{$^{1}$Deutsches Elektronen-Synchrotron DESY, Hamburg, Germany\\
$^{2}$University of Hamburg, Hamburg, Germany\\
$^{3}$University of Bonn, Bonn, Germany}
\def\corrAuthor{Paul Schütze}

\def\corrEmail{\href{mailto:paul.schuetze@desy.de}{paul.schuetze@desy.de}}

\begin{document}
\onecolumn
\firstpage{1}

\title[electronCT - An Imaging Technique Using VHEE]{electronCT - An Imaging Technique Using Very-high Energy Electrons} 

\author[\firstAuthorLast ]{\Authors}
\address{}
\correspondance{}

\extraAuth{}

\maketitle

\begin{abstract}
The electronCT technique is an imaging method based on the multiple Coulomb scattering of relativistic electrons and has potential applications in medical and industrial imaging.
It utilizes a pencil beam of electrons in the very high energy electron (VHEE, $\SI{50}{}$-$\SI{250}{MeV}$) range and a single detection layer for the determination of the beam profile.
The technique constitutes a projectional, two-dimensional imaging method and thus also qualifies for the tomographic reconstruction of samples.
Given the simplicity of the technical setup and its location behind the sample, the electronCT technique has potential synergies with VHEE radiotherapy, making use of the same electron source for both treatment and diagnostics and thus being a candidate for in-situ imaging and patient localization.
At the same time, several technical challenges arise from the measurement technique when applied for the imaging of living beings.

Measurements performed at the ARES linear particle accelerator at an electron energy of $\SI{155}{MeV}$ using a mouse phantom and a Timepix3 silicon pixel detector assembly demonstrate the feasibility of this technique.
Both projectional and tomographic reconstructions are presented and the potential and limits of the technology are discussed.

\tiny
 \keyFont{ \section{Keywords:} electronCT, medical imaging, multiple scattering, ARES, VHEE, Timepix3, radiation therapy} 
\end{abstract}

\section{Introduction}

In the past years, the field of radiation therapy has seen fast developments, driven by particle accelerator technologies that leverage the use of the very high energy electron (VHEE) regime, which typically refers to electrons in the energy range of $\SI{50}{MeV}$ to $\SI{250}{MeV}$, and the advancements in FLASH radiotherapy~\cite{ref:VHEE, ref:FLASH1, ref:FLASH2, ref:FLASH3}. 
In order to achieve a safe and effective treatment in radiation therapy, reliable imaging strategies are inevitable.
The application of imaging techniques in the planning and performance of radiation therapy is condensed in the term of image guided radiation therapy (IGRT).
A large number of imaging methods can be and are applied in the context of IGRT, such as X-ray imaging and CT, magnetic resonance imaging (MRI), positron emission tomography (PET), ultrasound imaging and camera-based imaging~\cite{ref:IGRT}.
While a good accuracy of a few millimeters can be achieved in aligning the treatment beam with an image using separate devices for these tasks, a unification of the treatment and an imaging device could improve this accuracy significantly and has been achieved for photon therapy~\cite{ref:tomotherapy}.

We present an alternative imaging technique called electron computed tomography (electronCT), that is based on highly energetic charged particles such as VHEE and their interaction with matter, in particular the multiple Coulomb scattering of the particles traversing the patient.
This measurement technique enables the use of the same accelerating structure for both treatment and diagnostics in the context of radiotherapy with VHEE and thus naturally creates a common coordinate system and does not require any alteration of the instrumentation around the patient.
This makes electronCT a candidate for an IGRT imaging technique in VHEE radiotherapy, for example for the in-situ localization of tumors or as an input for the alignment of high-resolution X-ray, CT or MRI images with the coordinate system applied for the treatment.

This article presents the concepts of the electronCT technique and shows proof-of-concept measurements.
Furthermore, the potential and limits of this method are discussed.

\section{The electronCT Technique} 

The technique of electronCT relies on the multiple Coulomb scattering of highly energetic particles in matter.
When traversing matter, charged particles, typically of momenta in the order of few hundreds of $\SI{}{\MeV}$, are stochastically deflected by the electrostatic force of the material's nuclei, leading to an effective deflection when traversing an object.
The effective deflection depends on the radiation length $X_0$ and the thickness $l$ of the material, as well as on the charge and momentum $p$ of the incident particle.
This scattering process and with it the effective deflection angle distribution are theoretically described by Moli\`ere's theory~\cite{ref:Moliere, ref:MoliereBethe}.
The central part of this distribution of deflection angles can be described by a Gaussian distribution centered around zero.
The width $\theta_0$ of this distribution is commonly approximated via a formula introduced by Highland~\cite{ref:Highland}, with the parameters revised by Lynch\&Dahl~\cite{ref:LynchDahl}:
\begin{equation}
\theta_0 = \frac{\SI{13.6}{\MeV}}{\beta c p} z \sqrt{\frac{l}{X_0}} \left( 1 + 0.038 \ln\left( \frac{l}{X_0} \right) \right),
\label{eq:Highland}
\end{equation}
with $\beta$ as the velocity in fractions of the speed of light $c$ and $z$ the charge number of the incident particle.
The width of the distribution depends on the radiation length of the material traversed, increases with the thickness of the material and is reduced for higher momenta.

In the electronCT technique, the amount of material traversed by a beam is measured by determining the opening angle of a collimated beam of particles after traversing a sample.
The amount of material is defined as the material's thickness normalized to its radiation length, $\epsilon=l/X_0$, and often referred to as the material budget.
This measurement can be accomplished by detecting the transverse beam profile using a single detection layer downstream of the sample, employing position sensitive radiation detectors.
Assuming an incident beam of charged particles with low transverse size and low divergence, the width of the transverse beam profile at a given distance behind the sample is thus a measure for the material budget along the path of the particles.

While many types of radiation detectors are applicable and should be chosen depending on the expected size and intensity of the beam delivered by the particle accelerator, this publication presents the use of silicon pixel detectors which are typically applied in high energy physics.
Silicon detectors measure the amount of energy deposited in a sensitive volume via ionisation processes and a segmentation of the sensor in pixels allows for the retrieval of two-dimensional information on a particle's traversal position~\cite{ref:semiconductorLutz}.
They are typically optimized for the tracking of individual particles through a multi-detector setup, thus their dynamic range is optimized for a small number of particles per readout cell.
In this case, silicon pixel detectors come with the benefit of less signal per readout channel and thus a lower required dynamic range compared to sensors segmented in strips.

\begin{figure}
    \begin{center}
    \includegraphics[width = 0.7\linewidth]{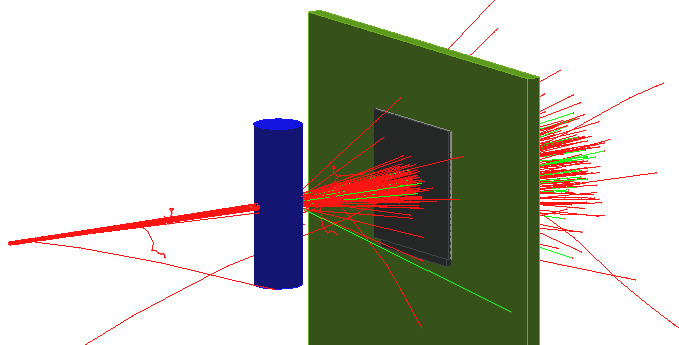}
    \caption{Visualisation of the simulation corresponding to the acquisition of a single data point acquired via the electronCT technique. A beam consisting of one hundred electrons with a width of $\SI{100}{\micro\meter}$ (red lines) is scattered at a sample (blue) and consecutively passes through a silicon detector (grey). Along the electron trajectories, a small number of Bremsstrahlung photons is generated (light green).}
    \label{fig:eCT_vis}
    \end{center}
\end{figure}

\begin{figure}
    \begin{center}
    \includegraphics[trim={0 0 0 79px},clip,width = 0.49\linewidth]{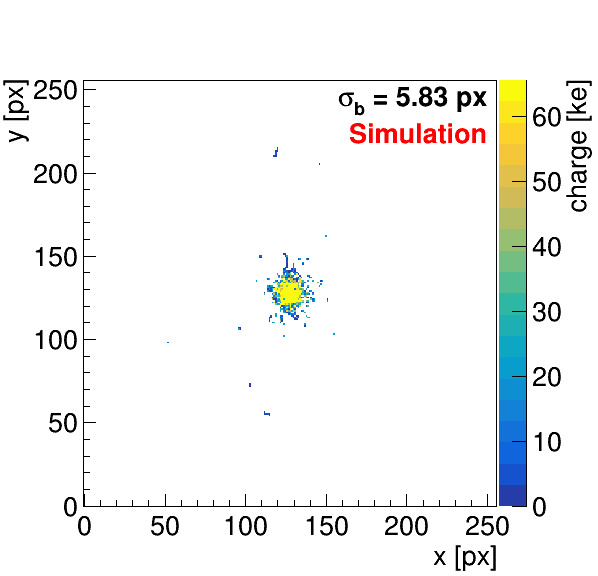}
    \includegraphics[trim={0 0 0 79px},clip,width = 0.49\linewidth]{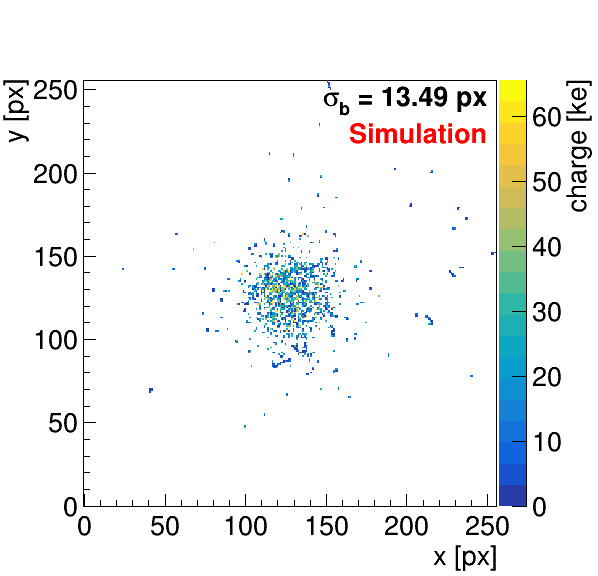}
    \caption{Simulation of beam profiles as detected by the silicon pixel detector in a simulation such as shown in Fig.~\ref{fig:eCT_vis} for the beam not passing through the sample (left) and traversing the center of the sample (right). Shown is the collected charge in units of kiloelectrons as a function of the impact pixel coordinates.}
    \label{fig:eCT_sim}
    \end{center}
\end{figure}

Figure~\ref{fig:eCT_vis} shows the visualisation of the simulated acquisition of a single data point with this technique, performed via the semiconductor detector simulation framework Allpix$^2$~\cite{ref:APSQ}, which utilizes the software toolkit Geant4~\cite{ref:G4} for the simulation of the interaction of particles with matter.
The incident beam, of which the individual particle trajectories are shown in red, is scattered at the sample (blue) before it is detected in the silicon detector (grey). In addition, a small number of Bremsstrahlung photons is generated (light green).
The sample in this case consists of two nested cylinders with radii in the range of small medical samples and radiation lengths in the range of tissues.
Figure~\ref{fig:eCT_sim} shows two examples for simulated beam profiles, with (right) and without sample present (left). Both beam profiles are displayed as the charge collected in the sensor, in units of kiloelectrons, as a function of the impact pixel coordinates and can clearly be distinguished from each other.
Gaussian distributions are fitted to the projections of the beam profiles onto the $x$- and the $y$-axes and the corresponding beam sizes are determined as widths of the fitted distributions in units of pixels. The means of these widths, $\sigma_b = \left(\sigma_x+\sigma_y\right)/2$, are indicated in the corresponding graphs in Figure~\ref{fig:eCT_sim}.
The simulation demonstrates that a measurement of the beam profile of an initially collimated electron beam via a pixelated silicon sensor is sensitive to the traversal through a sample prior to the detection.
In addition, this simulation was applied for designing the experimental layout.
For this simulation, Allpix Squared version 3.0.3~\cite{ref:APSQv300}, compiled with Geant4 version 11.2.1, has been used to simulate the interaction of a beam containing 1000 primary electrons with the sample and the detector, applying a Geant4 physics list with the electromagnetic constructor \textit{Livermore}.
It should be noted that for the visualisation in Fig.~\ref{fig:eCT_vis} a beam of only one hundred electrons has been simulated for a better visibility.


Two-dimensional or projectional imaging can be achieved by the bunched particle beam rastering the sample or by moving the sample across a fixed beam, such that each electron bunch and thus each measured width of a bunch profile can be attributed to a certain impact position on a virtual transverse plane through the sample.

Three-dimensional imaging is enabled by recording projections at different impact angles. This can be achieved by either rotating the sample or, similarly to X-ray based computed tomography (CT) measurements, the rotation of the particle source, here the accelerating structure, and detector around the sample. The rotation of the particle accelerator around the sample or patient is technologically available in a few modern radiation treatment facilities~\cite{ref:GantryHD, ref:GantrySummary}.

Consequently, the experimental setup comprises an accelerating structure delivering a beam with momenta of a few hundreds of $\SI{}{\MeV}$ and a beam size in the order of a few $\SI{100}{\micro\meter}$, and a silicon pixel detector capable of coping with a high data rate.

A similar technique, also based on multiple Coulomb scattering but applying the measurement of individual electron trajectories of $\SI{}{\GeV}$-electrons in a large beam, has proven to provide 3D imaging with resolutions in the order of $\SI{100}{\um}$ and good contrast-to-noise ratios for a wide range of material densities, at the downside of a low particle rate and thus extensive measurement times~\cite{ref:tbmst}.
The development of scanning the sample with a pencil beam brings the opportunity to drastically decrease the measurement time.
In addition, it has the potential to reduce the complexity of both the setup and the analysis by omitting the need to reconstruct individual particle trajectories and thus to reduce the time required for the reconstruction of an image.

\section{Experimental Setup}

\begin{figure}
    \begin{center}
    \includegraphics[trim={300px 0 400px 0}, clip, width = 0.9\linewidth]{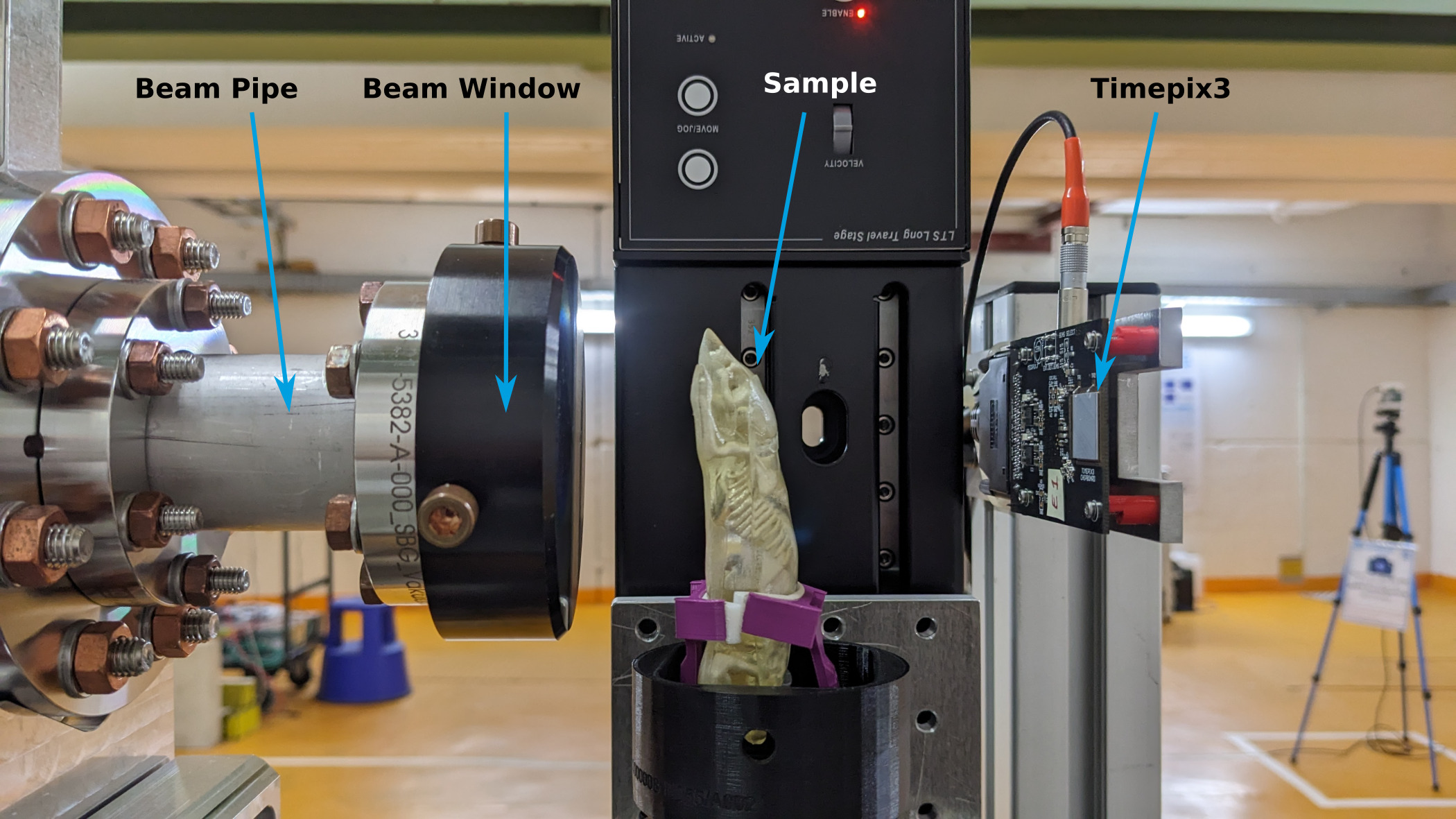}
    \caption{Picture of the electronCT setup at the ARES accelerator facility. The mouse phantom is mounted on a positioning stage and located in the center of the experiment. The beam exit window, secured by a black metallic cover, is to the left and the detector on the right hand side.}
    \label{fig:setupImage}
    \end{center}
\end{figure}

The measurements shown in this work were performed at the Accelerator Research Experiment at SINBAD (ARES)~\cite{ref:ARES} at DESY, Hamburg, using a Timepix3 silicon pixel detector assembly~\cite{ref:TPX3}.
A medical phantom resembling a mouse was placed between the electron extraction window of the accelerator and the detector.
The individual components were positioned as closely together as possible while assuring a clearance of the motion stages with respect to further components, leading to distances between the beam window and the phantom of $\SI{68}{mm}$ and between the beam window and the silicon detector of $\SI{134}{mm}$.
A picture of the main experimental setup can be seen in Figure~\ref{fig:setupImage}.

The measurements presented herein use a sample mounted to a 4D positioning system which rasters the sample across a fixed beam.

\subsection{ARES}

The ARES accelerator is a linear electron accelerator delivering ultra-short bunches with a charge of up to few hundreds of picocoulombs~\cite{ref:ARES}.
It is designed for a kinetic energy of up to $\SI{155}{\MeV}$~\cite{ref:ARES} and a bunch repetition rate of between $\SI{1}{\Hz}$ and $\SI{50}{\Hz}$.
For this experiment a repetition rate of $\SI{10}{\Hz}$ was applied and significantly lower bunch charges, estimated to be in the order of $\SI{1}{\femto\coulomb}$ and less, were achieved via the use of the dark current, a current generated by field emission in the accelerating structure~\cite{ref:DarkCurrent}.
The beam was transversely focussed onto the sample using the last quadrupole magnets of the accelerator beamline.
With this, a symmetric beam spot with an RMS size of around $\SI{320}{\um}$ was reached at the position of the sample and of about $\SI{360}{\um}$ at the detection plane, as will be presented below.
The transverse beam size was observed to be dominated by the scattering at the beam exit window, consisting of a titanium (Ti Grade 5) foil of $\SI{50\pm 5}{\um}$ thickness.

\subsection{Timepix3 Silicon Pixel Detector}

The choice of the detector is a crucial parameter for the electronCT method.
For the presented application, a large readout buffer and a large dynamic range are two main demands on the detection layer.
The former is motivated by the fact, that at the traversal of a single bunch a large area of the detector and thus a large quantity of pixels is hit within a small amount of time, and all buffered data has to be read out before the arrival of the succeeding bunch to avoid potential dead times.
The requirement for a large dynamic range arises from the fact, that the number of incident electrons remains constant, while the width of the beam profile strongly changes depending on the traversed material. In this experiment, both large peaks and long tails in the beam profile need to remain resolvable.

One type of detectors that fulfil these requirements are assemblies based on the Timepix3 or Medipix3 readout chips, which are already used in several medical applications~\cite{ref:TPX_MARS, ref:TPX_PET}.
Timepix3 readout chips feature an array of $256\times 256$ pixels with a pitch of $\SI{55}{\um}\times\SI{55}{\um}$ and provide charge information via Time-over-threshold (ToT) detection with a TDC resolution of 10 bit, as well as a time-of-arrival (ToA) measurement.
The assembly used herein consists of a Timepix3 readout chip bump-bonded to a silicon sensor with a thickness of $\SI{100}{\um}$.
The sensor was operated at a bias voltage of $-21\unit{V}$ to ensure full depletion~\cite{ref:W5_E2} and read out using a Katherine readout system~\cite{ref:Katherine} controlled via the TrackLab software framework~\cite{ref:Tracklab}.

\subsection{Phantom \& Positioning}

The phantom represents a mouse body with the tissue additively manufactured from resin, a skeleton made from a water-gypsum mixture and organs formed from agarose~\cite{ref:Berta}.
These materials have been chosen due to their manufacturing properties, low costs and their X-ray attenuation coefficients close to those of a real mouse.
It should be noted that the radiation length, as the underlying material parameter for electronCT measurements, does not scale linearly with the X-ray attenuation coefficient, but both strongly depend on the atomic number. However, the X-ray attenuation coefficient also exhibits a dependency on the particle energy, which is not the case for the radiation length.
Hence, although the materials contained in the sample are not expected to reproduce the exact multiple Coulomb scattering behaviour of a biological sample, it is expected to provide a good benchmark for electronCT measurements.

Linear translation stages of the type LTS300 by Thorlabs~\cite{ref:ThorlabsStages} were utilized for scanning the phantom across the beam in the two transverse dimensions, while the beam was kept at a constant position and the detector mounted on a fixed stand. The stages feature an accuracy of about $\SI{2}{\micro\meter}$. The sample can also be moved along the beam axis, but the range of motion in this dimension is severely restricted as the experiment setup is optimized for minimum distances along the beam axis.
The sample is furthermore mounted to a rotational stage of the type PRM1/MZ8 by Thorlabs\cite{ref:ThorlabsRotationStage} with a vertical rotation axis, featuring a sub-degree rotational precision. This allows for an illumination from different angles and thus a tomographic measurement.

\section{Methodology}

\subsection{Scan Sequence}

For two-dimensional measurements, the sample is moved across the $x$-$y$-plane, with the $x$-axis as the horizontal, and the $y$-axis as the vertical axis.
The scans have been conducted using a continuous motion along the $x$-axis at discrete steps along the $y$-axis, forming a serpentine path.
In this schematic, the minimum achievable resolution of the resulting images is given by the velocity of the motion stage along $x$ and the bunch repetition rate of $\SI{10}{Hz}$, and the step size along $y$.
With a motion velocity of e.g. $\SI{5}{\mm\per\s}$, the bunches sample the phantom with a spacing of $\Delta x = \SI{0.5}{\mm}$.

Three-dimensional, tomographic measurements via electronCT require a rotation of the sample, which is achieved by a rotation around the $y$-axis with the angle denoted as $\varphi$.
For these studies, the motion system was set up to perform two-dimensional scans as mentioned above for a configurable number of rotation angles consecutively.

An alternative sequence has been defined as a scan in the $x$-$\varphi$-space at a fixed $y$-position, representing a single horizontal line for several rotation angles, which allows for the tomographic reconstruction of a single slice of the sample through the $x$-$z$-plane. The advantage of this measurement in comparison with a full three-dimensional scan is the potential to study the tomographic reconstruction potential and performance at a drastically reduced measurement time.

It should be noted that for all scan types the sequence of data points taken is not relevant for the data analysis and can be optimized e.g. for reducing the measurement time.

\subsection{Data Acquisition}

For electronCT measurements, the electron bunch is focused at the detector, such that in most cases several electrons contribute to the signals of individual pixels.
The time structure of the bunch does not allow for a separate detection of the individual primary particles and hence the summed deposited charge below each pixel is measured.
As a consequence, a high bunch charge focused onto a small number of pixels would lead to a saturation of the detector front-end dynamic range and should thus be avoided.

The Timepix3 chip was configured with a threshold of about four kiloelectrons. It should be noted that the unit kiloelectrons ($\SI{}{ke^{-}}$) here denotes the number of charge carriers collected per pixel and not the number of primary electrons contributing to the signal.
The threshold corresponds to about $40\%$ of the signal induced by a single electron and is thus sufficient for the detection of individual beam particles.

The readout chip performs a zero-suppression, i.e. only pixels detecting a signal larger than the configured threshold are registered, and can be operated in two modes, the so-called \textit{sequential} (often called \textit{frame-based}) readout mode or the \textit{data-driven} mode.
In the frame-based mode a readout frame is defined by an external signal and the data from all pixels that registered a signal above the configurable threshold within this frame is read out after the frame has ended. The Katherine readout system allows to delay the frame start with respect to an external signal and to set a configurable frame duration. In this readout mode the system can be configured, such that each frame represents the signals from an individual bunch.
The data-driven readout mode represents a continuous detector readout, leading to a data set of all signals exceeding the threshold, containing the corresponding pixel addresses including timestamps. In this mode, the timestamps of the signals can be used to group them into signals arising from individual bunches during the post-processing of the data.

In case of the frame-based readout, frames have been triggered via the accelerator machine clock, with the delay manually adjusted, such that bunches arrive within the first microsecond of a frame. The frames were configured to feature a width of \SI{10}{\micro s}.
The data contain the coordinates and ToT values of all pixels registering a signal above the threshold per readout frame as well as their ToA within the frame.

In the data-driven readout mode, the data contain the coordinates and ToT values of all signals above the threshold as well as each signal's ToA within the data stream. While the global timestamp of the readout chip itself features a range of $\SI{409.6}{\micro s}$, the Katherine DAQ system is capable of extending the timestamp to a range of more than 20~days and is hence sufficient for all measurements performed.

\subsection{Data Analysis}

\subsubsection{Data Processing}

\begin{figure}
    \begin{center}
    \includegraphics[width = 0.95\linewidth]{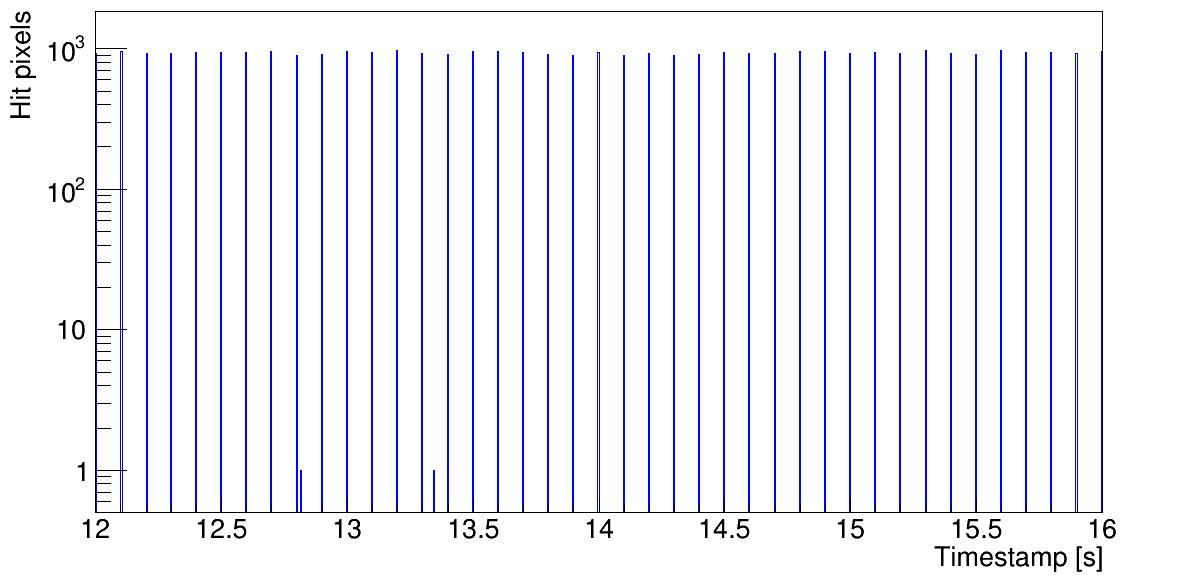}
    \caption{Time structure of the beam generated by ARES as recorded by the Timepix3 detector assembly, represented as the number of hit pixels as a function of time.}
    \label{fig:ares_timestructure}
    \end{center}
\end{figure}

The data were converted and every frame was interpreted using the software framework Corryvreckan~\cite{ref:corry}.
To interpret data recorded with the Katherine readout system via this framework, a module for reading data recorded by TrackLab was added.
In the case of measurements applying the data-driven readout mode, no inherent frame structure is available that would allow to assign the detector data to a given state of the motion system. Instead, a continuous array containing all pixels with a signal above the threshold with the corresponding ToA is obtained. An example of the time structure of the recorded data is shown in Figure~\ref{fig:ares_timestructure} by means of the number of hits as a function of their corresponding time stamps.
The time structure shows a clear, regular pattern caused by the ARES bunch structure leading to large numbers of hit pixels in a vanishingly small time frame with a distance of $\SI{100}{\milli\second}$ as expected from the bunch repetition rate of $\SI{10}{Hz}$.
Infrequent signals are recorded in between the arrivals of two bunches as seen at the positions of $\SI{12.8}{s}$ and $\SI{13.3}{s}$ and can be caused by noise or uncorrelated radiation.

The ToA information of each signal was used to group them into frames, where every frame represents the data induced by a single bunch.
This was achieved with a minimal bias on the actual event separation by an algorithm splitting the data stream at positions where no significant group of signals has been recorded.

\begin{figure}
    \begin{center}
    \includegraphics[trim={0 0 0 79px},clip,width = 0.49\linewidth]{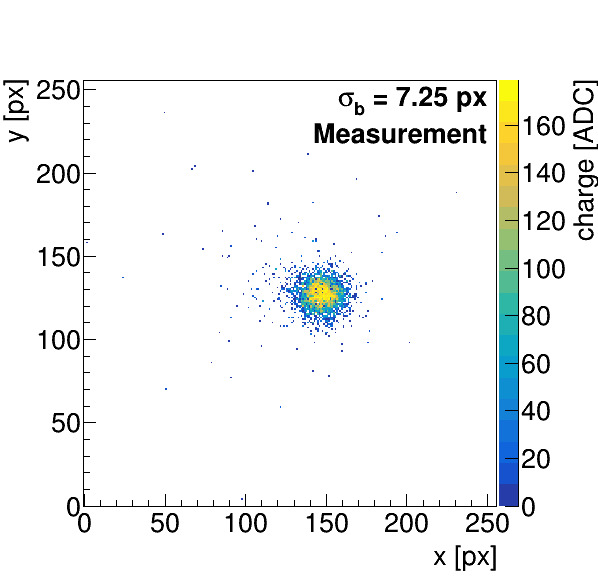}
    \includegraphics[trim={0 0 0 79px},clip,width = 0.49\linewidth]{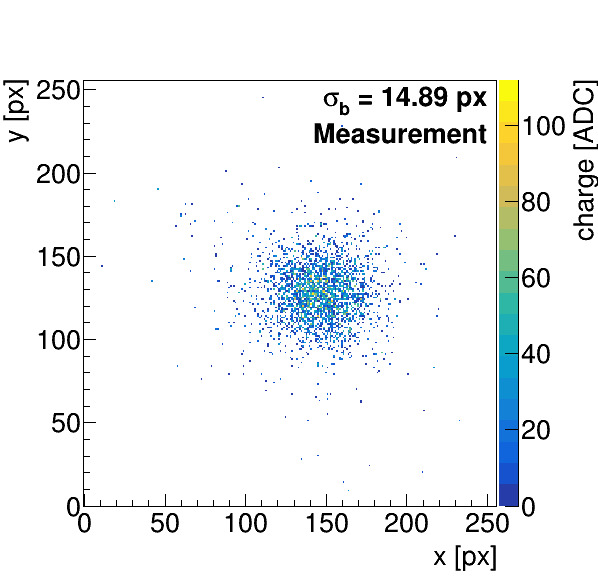}
    \caption{Charge map of the detector signal from two individual bunches, one not traversing the sample (left) and one traversing it (right).}
    \label{fig:chargeMaps}
    \end{center}
\end{figure}

Figure~\ref{fig:chargeMaps} shows two charge maps for individual frames, representing a mapping of all hits to their pixel coordinates with the ToT as color scale, one corresponding to a frame in which the beam does not pass through the sample (left) and one to a frame for which the sample was traversed by the beam (right).
It is readily visible that the multiple scattering of the beam particles at the object enlarges the beam laterally at the position of the sensor.
The sensor front-end saturates in case of large amounts of particles impinging in individual pixels, leading to a plateau visible in Figure~\ref{fig:chargeMaps} (left).

For the analysis of this data, a dedicated module in the Corryvreckan data analysis framework projects these maps onto the x- and the y-axes and performs a fit of a Gaussian distribution to each projection, retrieving the widths of the fitted distributions, $\sigma_x$ and $\sigma_y$.
Previous measurements showed that the scattering angles of individual particles along two orthogonal axes are uncorrelated, while the means of the distributions strongly correlate~\cite{ref:tbmst}. This suggests that $\sigma_x$ and $\sigma_y$ represent the same quantity and hence the measurements of the beam size along $x$ and $y$ can be treated as two independent measurements in order to reduce uncertainties. 
This procedure is mathematically equivalent to the generation of two images from the individual widths and averaging the image content.
In the following, the beam size per frame is determined as the mean of the standard deviations retrieved from both fits, $\sigma_b = \left(\sigma_x+\sigma_y\right)/2$.

\subsubsection{Image Generation}

The beam size per frame, as the measured quantity of interest, is attributed to the positions of the linear motion stages and, in case of tomographic measurements, the rotation stage. 
For this, the current stage positions are queried at a $\SI{10}{Hz}$ frequency and stored with their corresponding UNIX time stamps. As this process is not synchronized to the DAQ system, a synchronisation has to be performed in the post-processing and is implemented as follows: 
In a first step, the position of an absorber with a sharp edge, located next to the sample, is determined. This is achieved by means of a slow, manually controlled motion of the stage, in which the beam samples the edge of the absorber, and the simultaneous observation of a change in the charge maps recorded by the detector.
Subsequently, before the beginning of each two- or three-dimensional scan, the absorber is moved into and out of the beam path as a part of the automated scan sequence.
With the position of the absorber edge known, the exact time of the edge transition can be located both in the data stream of stage positions and in the beam profile data, as the latter exhibits a drastic increase in deposited charge at the time of this transition.
The two data streams are then correlated using their individual time stamps corrected by the determined offset. No significant drift between the time stamps has been observed.
This synchronization strategy requires stable beam conditions in terms of intensity and beam position, which was measured to be satisfied with relative intensity fluctuations in the order $\SI{5}{\percent}$ and variations of the beam position of less than $\SI{20}{\micro\meter}$ at the position of the detection plane.

With the synchronized data streams of stage positions and beam profile data, each reconstructed beam size value can be attributed to a certain point in the $x$-$y$ (2D), $x$-$y$-$\varphi$ (3D) or $x$-$\varphi$ (Single Slice) scan range. 
Equally spaced image cells within the scan range can be defined such, that either a single value is obtained per image cell, or the average of multiple values falling into a cell's range is determined. 
The former enables a better image resolution, while the latter has the potential to improve the image contrast.

In case of three-dimensional measurements, the beam size obtained by a measurement as discussed above is corrected by subtracting a background value, determined as the mean observed beam width within a region where the beam did not pass through the sample.
This correction compensates for the finite width of the beam after traversing the beam window and takes into account the effect of multiple Coulomb scattering in air.

Subsequently, the data set is structured in sinograms, which are representations of the corrected beam size as a function of the $x$- and $\varphi$-positions for individual steps along the $y$-axis, using the assignment of beam size values to the $x$-$y$-$\varphi$ phase space of the motion system.
From these sinograms, slices of the material through the $x$-$z$-plane are obtained via inverse Radon transforms~\cite{ref:DeansRadon}, computed individually for every step along $y$.
For the measurements presented here, the open source software framework scikit-image~\cite{ref:scikitWebpage, ref:scikitArticle} was used to perform a filtered back-projection applying a ramp filter for the reconstruction of individual slices.
Subsequently, a wavelet denoising algorithm included in the framework following~\cite{ref:denoise} is applied on the final images for an improvement of the image quality.

\subsection{Beam characterization}

The lateral size of the electron beam generated by the ARES linear accelerator was characterized in terms of a measurement as a function of the longitudinal distance from the beam exit window.
For this measurement, the Timepix3 detector assembly, read out via the data acquisition system described above, was mounted to the three-dimensional $x$-$y$-$z$ linear motion stage system and a scan along the beam axis was performed covering a distance range from the beam exit window of $\SI{27}{mm}<z<\SI{317}{mm}$ with a step size of $\SI{10}{mm}$.
The minimum distance was limited by mechanical constraints and the maximum distance given by the range of the motion stage.
Per step, the stage positions were kept constant for $\SI{60}{s}$.

\begin{figure}
    \begin{center}
    \includegraphics[trim={0 0 40px 40px},clip,width=0.65\linewidth]{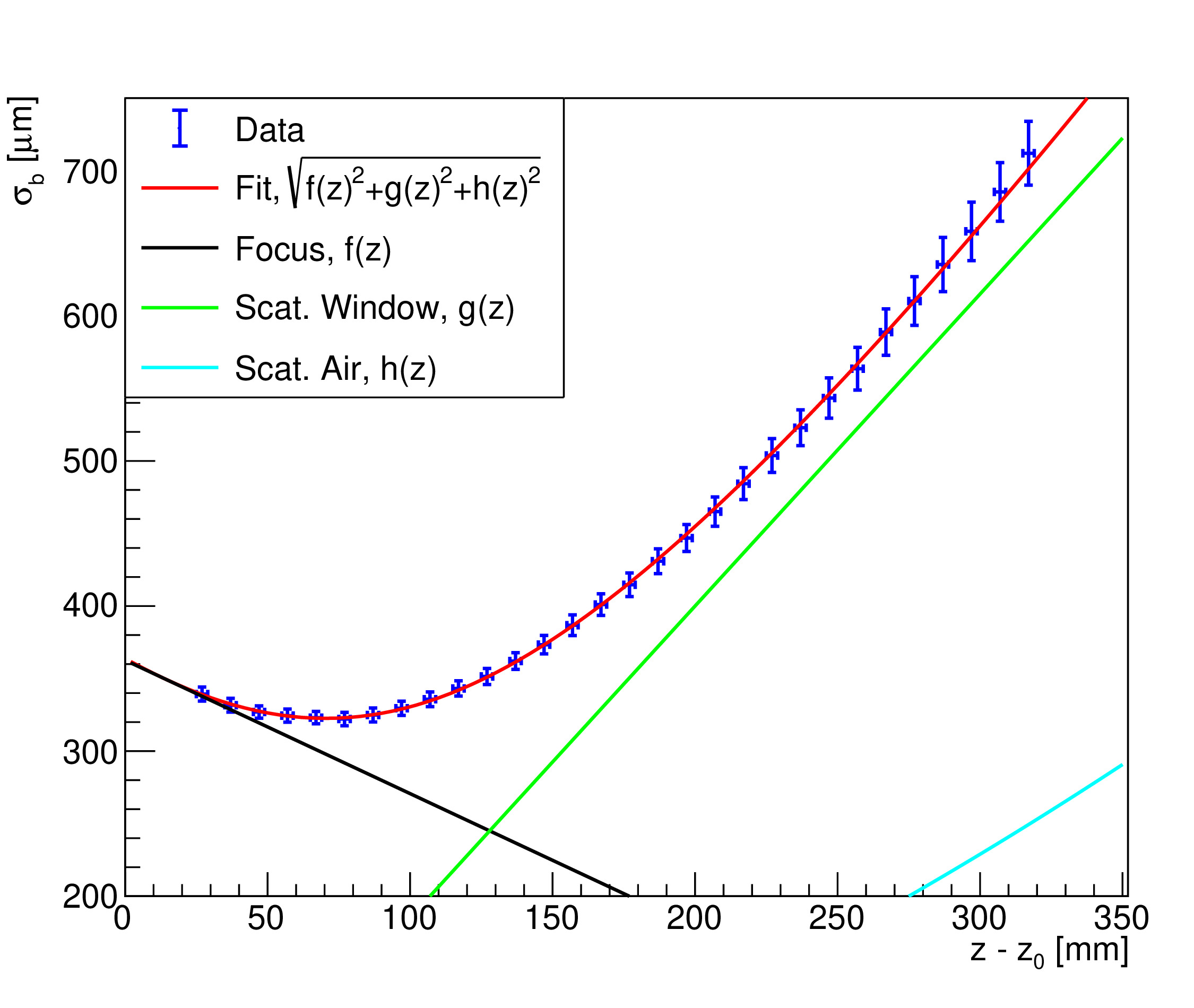}
    \caption{Measurement of the beam width $\sigma_b$ as a function of the detector distance from the beam exit window. A fit to the data represents a quadratic superposition of a focusing and two scattering contributions from the beam exit window and air.}
    \label{fig:zscan}
    \end{center}
\end{figure}

The beam size was evaluated on a per-bunch basis and the mean beam size per step is displayed in Figure~\ref{fig:zscan}, with the error bars representing the RMS of the beam sizes per step.

The beam features a focal point at a distance of about $\SI{71}{mm}$ from the beam exit window with a beam size of $\SI{322}{\micro\meter}$ as a result of focusing the beam onto the sample.
The trend as a function of the longitudinal distance can qualitatively be described via three terms added quadratically:
the contributions represent a focusing of the beam via the quadrupoles ($f(z)$), the scattering at the beam window ($g(z)$) (both polynomials of first degree) and a contribution describing scattering in air ($h(z)$).
The latter applies Equation~\ref{eq:Highland} for computing the RMS lateral displacement after Equation (34.20) of~\cite{ref:PDG-2022}.
This is indicated by a fit to the function
\begin{equation}
h(z) = \sqrt{f(z)^2+g(z)^2+h(z)^2} = \sqrt{(b_f+m_fx)^2+(b_g+m_gx)^2+\frac{a}{\sqrt{3}}z\theta_0(z)}
\end{equation}
with the free parameters $b_{f,g}$, $m_{f,g}$ and $a$.
The individual contributions are displayed in black, green and cyan.
A quantitative analysis was not performed due to systematic effects such as a saturation of the detector for small beam sizes and the limited data range.

The data shows that the beam is focused at the position of the sample and the rotation axis, located at a distance of $\SI{68}{mm}$ from the beam exit window, which enables an optimal image resolution.
The detection plane was positioned as close as possible, at a position of $\SI{134}{mm}$ with a beam size of $\SI{360}{\micro\meter}$.

\section{Results}

\subsection{Projectional Measurements}

In the following results, the beam size is displayed as a function of the sample position.
This beam size serves as an estimator for the traversed material budget, hence a two-dimensional measurement qualitatively represents a two-dimensional projection of the material budget onto the image plane.

\begin{figure}
    \begin{center}
    \includegraphics[trim={0 70px 0 70px},clip,width = 0.48\linewidth]{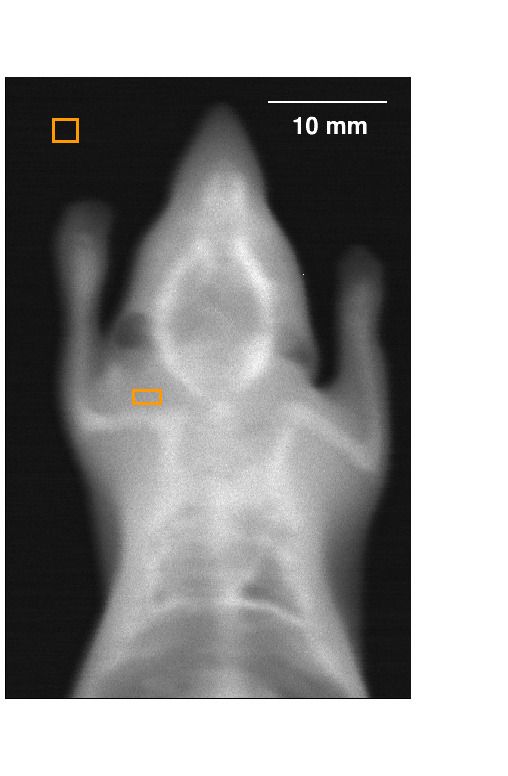}
    \includegraphics[trim={0 70px 0 70px},clip,width = 0.48\linewidth]{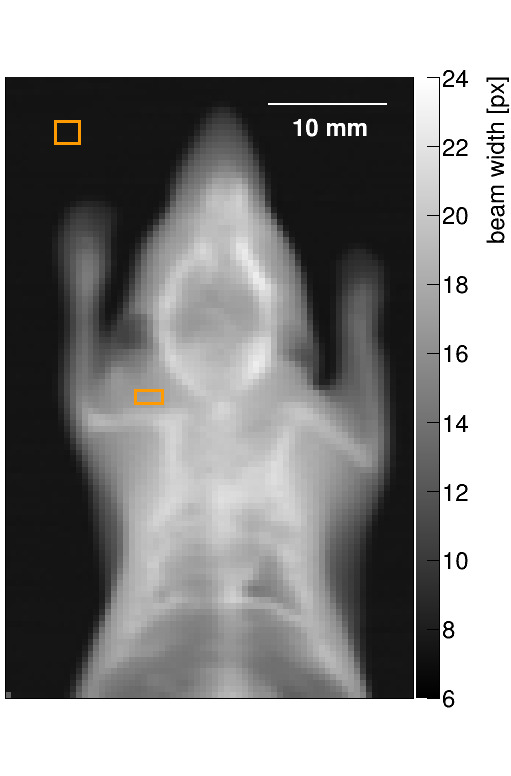}
    \caption{Two-dimensional electronCT measurements of a mouse phantom, evaluated with image cell sizes of $\SI{0.05}{mm}\times\SI{0.1}{mm}$ (left) and $\SI{0.5}{mm}\times\SI{0.5}{mm}$ (right). Orange boxes indicate the regions used for the determination of the image contrast.}
    \label{fig:result2D_Berta}
    \end{center}
\end{figure}

Figure~\ref{fig:result2D_Berta} shows the result of a measurement performed in the frame-based readout mode with a scan velocity of $\SI{0.5}{mm\per s}$ along the $x$-axis and 520 steps of $\SI{0.1}{mm}$ each along the $y$-axis.
The measurement time amounted to $\SI{620}{\min}$.

The data were evaluated with image cell sizes of $\SI{0.05}{mm}\times\SI{0.1}{mm}$ (left) and $\SI{0.5}{mm}\times\SI{0.5}{mm}$ (right).
Both images reveal many details of the sample: the skeleton can clearly be distinguished from the tissue and is resolved to a good level of detail, exhibiting the spine, the ribs, the arms and the skull.
The tissue can be well distinguished from the background.
The smallest features resolved in these images are the ribs with a thickness of about $\SI{0.5}{mm}$, which provides an upper limit for the achieved resolution.

It is readily visible, that while achieving a better resolution in Figure~\ref{fig:result2D_Berta} (left), Figure~\ref{fig:result2D_Berta} (right) exhibits less noise and thus a higher contrast.
The contrast-to-noise ratios (CNR) are determined by

\begin{equation}
\textnormal{CNR} = \frac{\mu_{\textnormal{sig}}-\mu_{\textnormal{bg}}}{\sqrt{\sigma_{\textnormal{sig}}^2+\sigma_{\textnormal{bg}}^2}}
\label{eq:CNR}
\end{equation}

with $\mu$ and $\sigma$ as the mean and standard deviation of values defined in a signal and a noise area.
As a signal region, a homogeneous region in the shoulder of the sample representing tissue, and as a noise region, an equally sized region outside the sample have been used and are indicated as orange boxes.
The CNRs result in 18.8 for the left image in Figure~\ref{fig:result2D_Berta} and 34.7 for the right image in Figure~\ref{fig:result2D_Berta}.

\subsection{Tomographic Measurements}

Two types of tomographic measurements have been performed using the presented phantom:
\begin{enumerate}
    \item $x$-$y$-$\varphi$ scan for a three-dimensional tomographic reconstruction.
    \item $x$-$\varphi$ scan for the tomographic reconstruction of a single slice in the $x$-$z$-plane.
\end{enumerate}

For measurement~1, a scan velocity of $\SI{1.5}{mm\per s}$ along the $x$-axis and 50 steps of $\SI{1}{mm}$ along the $y$-axis have been selected, with 37 projections recorded for a half turn, resulting in angular steps of $\SI{5}{\degree}$.
The measurements were performed using the Katherine readout system in the data-driven readout mode, recording a continuous data stream for around \SI{18}{\hour}.

\begin{figure}
\begin{center}
    \includegraphics[trim={0, 70px, 120px, 70px}, clip, width=0.22\textwidth]{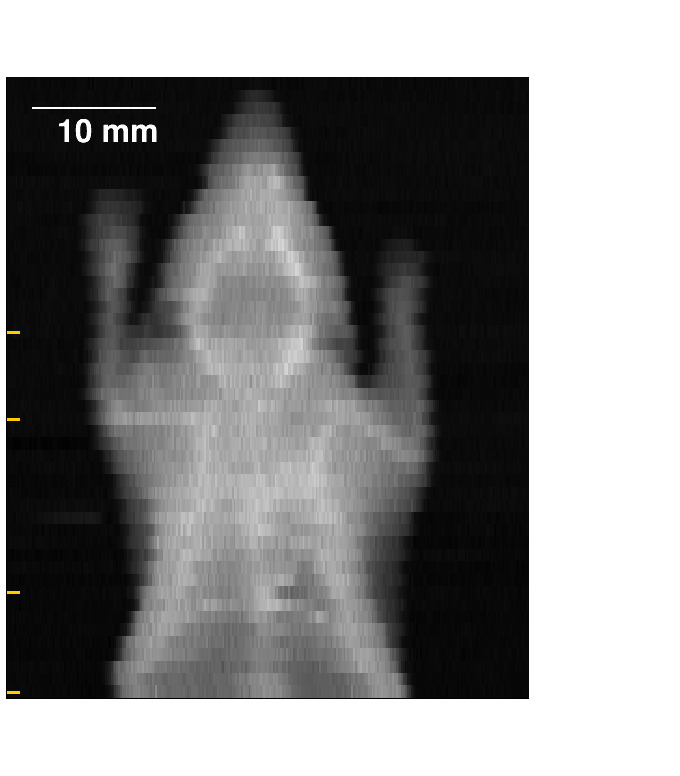}
    \includegraphics[trim={0, 70px, 120px, 70px}, clip, width=0.22\textwidth]{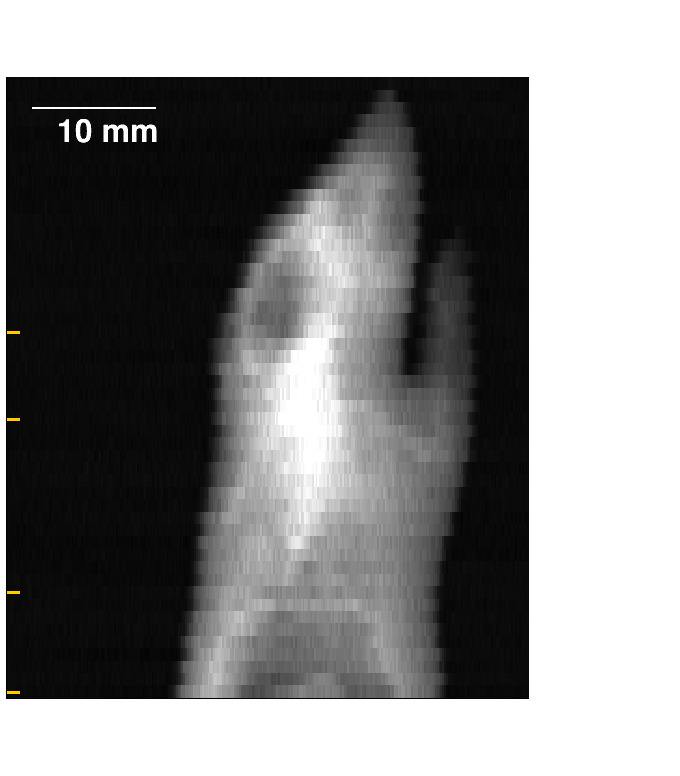}
    \includegraphics[trim={0, 70px, 120px, 70px}, clip, width=0.22\textwidth]{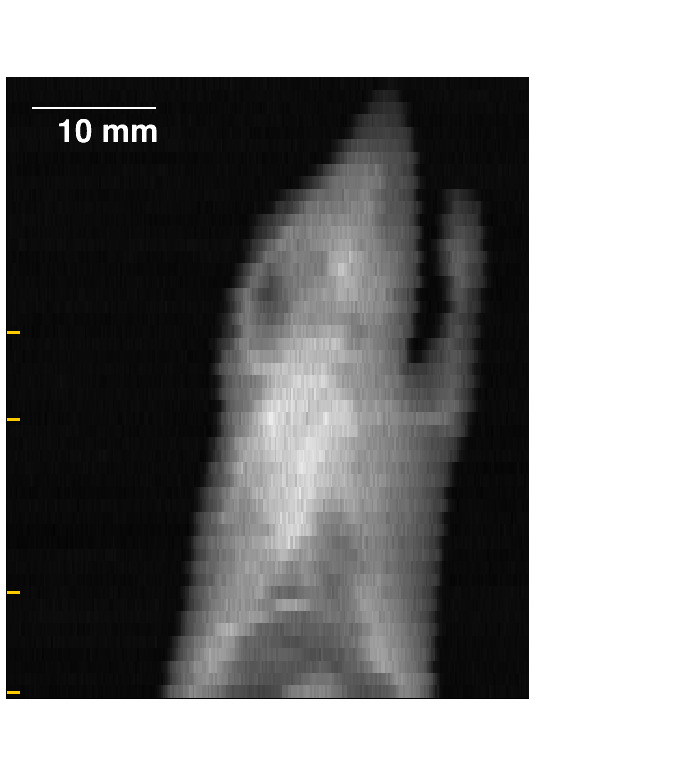}
    \includegraphics[trim={0, 70px, 120px, 70px}, clip, width=0.22\textwidth]{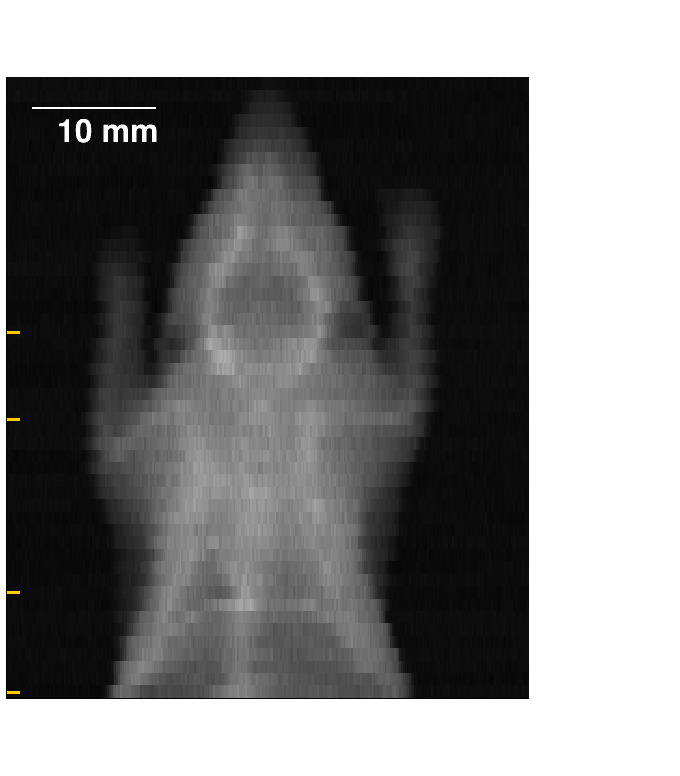}
    \includegraphics[trim={0, 70px, 0, 67px}, clip, height=0.262\textwidth]{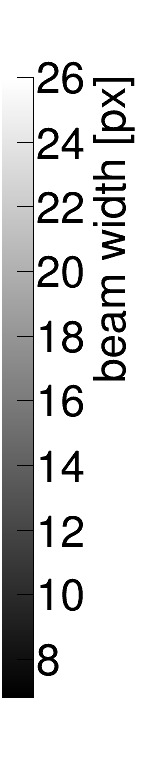}
    \caption{Two-dimensional projections of the phantom recorded for a tomographic measurement at rotation angles of $\varphi = \left\{ \SI{0}{\degree}, \SI{60}{\degree}, \SI{120}{\degree}, \SI{175}{\degree} \right\}$ (f.l.t.r.). Orange ticks on the left side of the images indicate the vertical positions for which reconstructions are presented.}
    \label{fig:run142_2Dprojection}
\end{center}
\end{figure}

From the three-dimensional scan, two-dimensional projection images can be generated for different rotation angles of the sample.
Figure~\ref{fig:run142_2Dprojection} shows four such projections for rotation angles of $\varphi = \left\{ \SI{0}{\degree}, \SI{60}{\degree}, \SI{120}{\degree}, \SI{175}{\degree} \right\}$ with image cell sizes of $\SI{0.15}{mm}\times\SI{1}{mm}$.
Due to the higher scan velocity and larger step size along $y$ with respect to the images presented in Figure~\ref{fig:result2D_Berta}, the resolution is inferior to these.
The projections acquired at rotation angles of $\SI{0}{\degree}$ (left) and $\SI{175}{\degree}$ (right) show similar features, but a reduced contrast can be observed in the latter. The reason for this is the positioning of the sample slightly off the rotation axis and thus the sample having different distances from the detector, which comes with an impact on the measured beam size.

\begin{figure}
    \begin{center}
    \includegraphics[width = 0.49\linewidth]{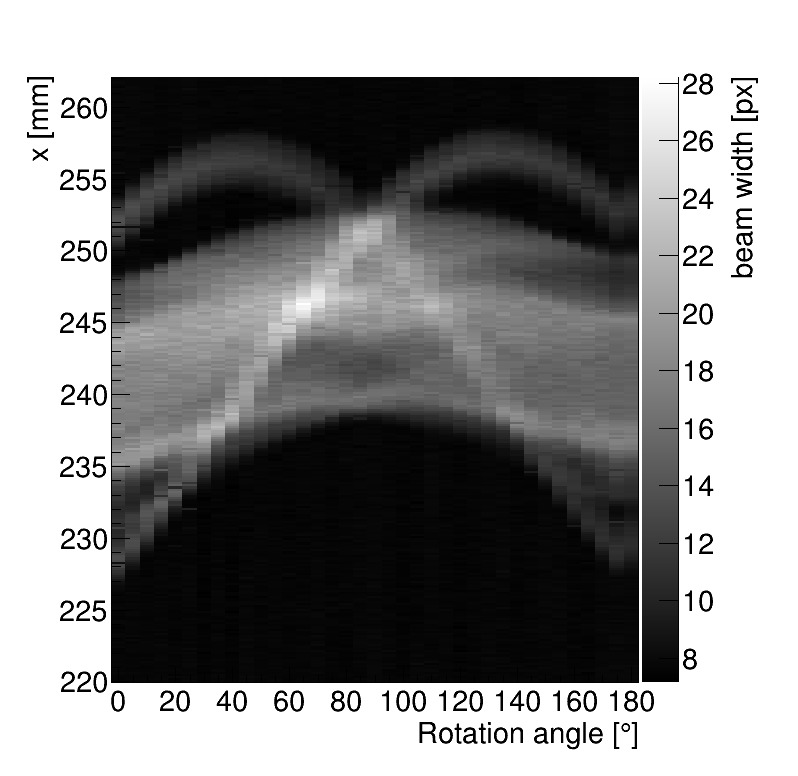}
    \caption{Sinogram generated from two-dimensional projections, such as shown in Figure~\ref{fig:run142_2Dprojection} for a vertical position representing the head of the phantom.}
    \label{fig:run142_sinogram}
    \end{center}
\end{figure}

Sinograms are generated for each horizontal line, hence for every scan step along the $y$-axis, combining the data of all rotation steps.
An example is shown in Figure~\ref{fig:run142_sinogram}, displaying the beam size as a function of $x$ and $\varphi$ for a vertical position inside the head region of the phantom.

\begin{figure}
\begin{center}
    \includegraphics[width=0.22\textwidth]{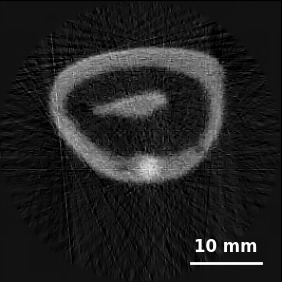}
    \includegraphics[width=0.22\textwidth]{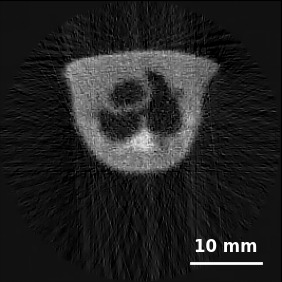}
    \includegraphics[width=0.22\textwidth]{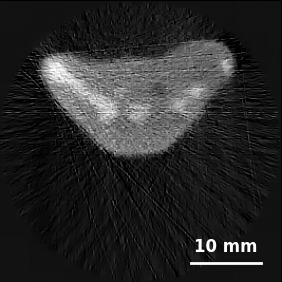}
    \includegraphics[width=0.22\textwidth]{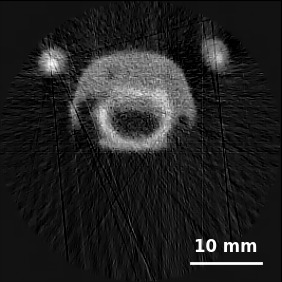}
    \includegraphics[height=0.228\textwidth]{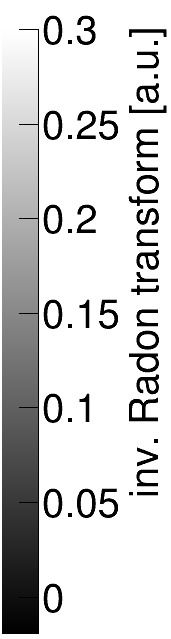}
    \caption{Three-dimensional electronCT measurement of a mouse phantom with an image cell size of $\SI{0.15}{mm}\times\SI{0.15}{mm}$, evaluated at four positions along the vertical axis representing different anatomical regions of the phantom.}
    \label{fig:result3D}
\end{center}
\end{figure}

Inverse Radon transforms are performed separately for each position along the $y$-axis, resulting in horizontal cuts, or slices, through the sample.
Figure~\ref{fig:result3D} shows four such slices for the vertical positions indicated via orange ticks on the left hand sides of the images shown in Figure~\ref{fig:run142_2Dprojection}.
They comprise regions of the abdomen including the spine (left), the lung including the spine and a cavity for an insertable heart (center-left), the upper arm and the shoulder (center-right) and the head including the skull and the arms (right).
The latter corresponds to the sinogram shown in Figure~\ref{fig:run142_sinogram}.
Despite a larger noise contribution compared to the two-dimensional scans, the features of the phantom can be recognized and a differentiation between air, tissue-like and bone-like material is possible.

Measurement~2 was performed at a lower scan velocity of $\SI{0.7}{mm\per s}$ along the $x$-axis, with 101 rotations covering a half turn.
This measurement is performed at a constant $y$-position within the head (cf. Figs~\ref{fig:run142_sinogram} and~\ref{fig:result3D} (right)) and thus yields a single horizontal cut through the sample.
The measurement time was \SI{101}{\minute}.

\begin{figure}
    \begin{center}
    \includegraphics[width = 0.4\textwidth]{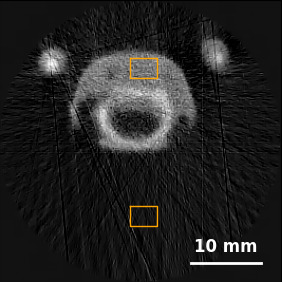}
    \includegraphics[width = 0.4\linewidth]{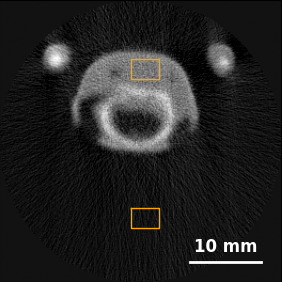}
    \includegraphics[height=0.413\textwidth]{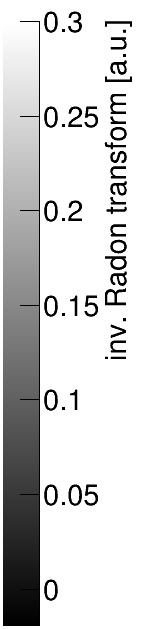}
    \caption{Comparison of two tomographic reconstructions of the sample, both acquired and evaluated at the same vertical position and displayed with image cell sizes of $\SI{0.15}{mm}\times\SI{0.15}{mm}$. (left) shows the enlarged result of an $x$-$y$-$\varphi$ scan shown in Figure~\ref{fig:result3D}. (right) shows the result of an $x$-$\varphi$-scan with increased amount of data. Orange boxes indicate regions used for the determination of the image contrast.}
    \label{fig:resultSlice}
    \end{center}
\end{figure}

The result is presented in Figure~\ref{fig:resultSlice} (right) and compared to a reconstruction slice at the same position from measurement~1 (left), both with image cell sizes of $\SI{0.15}{mm}\times\SI{0.15}{mm}$.
A clear improvement from (left) to (right) is visible in terms of contrast and resolution, as well as a reduction of artefacts.
The CNRs, calculated from signal and background regions indicated via orange boxes and determined via Equation~\ref{eq:CNR}, amount to 1.96 (Figure~\ref{fig:resultSlice}, left) and 4.62 (right).
These improvements are achieved by an increase in statistics: while Figure~\ref{fig:resultSlice} (left) contains the information of $10360$ frames, (right) contains information of $57700$ frames due to the reduced stage velocity and the increased number of projections.
Differences caused by a variation in the scanning procedure are neither expected nor observed.

It should be noted that the reconstruction via an inverse Radon transform assumes a linear dependency of the measured observable, here the beam width, on the quantity to be reconstructed, here the material budget density.
While Equation~\ref{eq:Highland} suggests a dependency of the opening angle and thus the beam width on the square root of the material budget, it has been found that using the detected beam width with a background subtraction as an input to the filtered backprojection, yields reasonable results with minor artefacts.
Further corrections mitigating non-linearities are expected to improve the image quality and are subject to current studies.

\section{Potential \& Limitations}

The presented measurements demonstrate the technological feasibility of the electronCT imaging technique and show reasonable resolution and contrast for macroscopic objects of sizes in the order of a few tens of millimeters, resolving details such as the skeleton of a mouse.
In the following, the potential and the limitations of this technology are discussed.

\subsection*{Scanning Strategy}

For the measurements presented above, a constant transverse position was chosen for the electron beam, while the phantom was moved across the transverse plane.
For the studies on phantoms such a strategy is applicable whereas it would be unfeasible for imaging in medical scenarios.

Instead, dipole magnets used for steering the transverse beam trajectory could be used to move the beam relatively to a static sample.
This however comes with the necessity to either move the detector synchronously with the center of the electron beam or to use a much larger detector that would be required to cover the full scan range of incident beam positions, and with the requirement of a small energy spread to avoid dispersion effects.

\subsection*{Spatial Resolution}

The achievable resolution for the electronCT technique depends on several parameters:
\begin{itemize}
    \item Step size: naturally the resolution of an image depends on the pitch of the individual image cells, which for this technique is limited by the distance between the impact positions of consecutive bunches or the step size along discretely scanned axes. This parameter also has a direct impact on the measurement time.
    \item Beam size: the transverse size of the beam can pose a limit on the spatial resolution of the obtained image, as individual electrons are scattered at different positions at the sample. This property is limited by the accelerator performance and the geometry of the experiment.
    \item Beam widening: already while traversing the sample, the beam size increases due to multiple Coulomb scattering. Hence, the spatial resolution suffers in the case of samples with a large thickness and high material budget.
    \item Beam stability: shot-to-shot fluctuations or long-term drifts of the beam position and intensity at the sample can affect the spatial resolution but could partially be recovered by correcting individual image points for the center of the detected beam profile.
\end{itemize}
For the measurements shown above, the limits on the spatial resolution were posed by the RMS beam size of around $\SI{320}{\micro\meter}$ for some of the measurements, while others were dominated by the step size. 
The phantom was chosen sufficiently small to not expect dominating effects from a beam widening inside the sample and the position of the beam center was found to be stable within less than $\SI{20}{\micro\meter}$.

\subsection*{Measurement Time}

A great challenge for this imaging technique lies in the measurement time.
The measurements presented above range from $\SI{101}{\minute}$ to $\SI{18}{\hour}$ and are thus unacceptable for medical imaging of living beings.
The measurement time however strongly depends on, and is in these studies limited by, the repetition rate of the accelerator in use.
Using higher repetition rates directly decreases the measurement time, such that images such as the ones shown in Figure~\ref{fig:result2D_Berta} with a repetition rate of e.g. $\SI{1}{kHz}$ could be performed within less than $\SI{10}{\minute}$.
For the detector in use, the readout bandwidth is limited to $\SI{85}{MHits\per\second}$, such that with less than 2000 hit pixels per frame (see Fig.~\ref{fig:ares_timestructure}), even repetition rates of up to $\SI{42}{kHz}$ and thus measurement times of less than a minute would be theoretically possible.
However, at very high rates, the synchronisation of the data acquisition with the scan of the beam position across the sample via dipole magnets requires a high precision for a proper image reconstruction.

At the same time, the requirement on the spatial resolution of an image depends on its purpose and could thus be much looser than the resolution achieved with the above studies.
As the resolution impacts the number of data points and thus the total number of electron bunches required for an image, imaging could be performed much faster by adapting the target resolution.

\section{Conclusions \& Outlook}

The feasibility of the imaging technique electronCT has been demonstrated by means of two- and three-dimensional measurements.
The method is based on the determination of the beam profile of a low-charge electron beam with energies in the range of few hundreds of $\SI{}{MeV}$ after the traversal of a phantom.
Projectional images as well as tomographic reconstructions of a medical mouse phantom have been acquired using the ARES linear accelerator for beam generation and a Timepix3 detector assembly as the detection layer.
The images exhibit many details of the phantom and show good resolution and contrast.
This proof of concept enables studies toward an application in the context of radiation treatment with VHEE, where this method could create synergies in applying the same accelerating structure for treatment and imaging.

The presented studies expose limitations and technical challenges of the technique in the prospect of medical imaging, which lie in the measurement time and with it artifacts arising from the motion of living beings, and the spatial resolution when applied to larger objects.
Hence, further studies on this technique are required to gain an understanding on the improvement potential of the measurement time, but also for an estimation of the dose delivered to a sample or patient in various imaging scenarios to explore its potential as imaging modality in a medical context.
In summary, electronCT represents a candidate for in-situ imaging in the context of VHEE radiotherapy and could contribute to IGRT, among others for the patient and tumor localization, under the premise of overcoming and gaining a further understanding of the aforementioned challenges.

Additional future studies serve the optimisation of this technique and comprise detailed simulations using the semiconductor detector simulation framework Allpix$^2$, which is capable of simulating the effects of multiple Coulomb scattering in a phantom as well as the detector response to the particle beam. Such simulations can serve as a guidance for an optimized measurement setup and strategy and provide insight into resolution limits and potential artefacts.
These limits will be studied via the imaging of more generic, geometric samples such as a Derenzo phantom~\cite{ref:Derenzo}.

As indicated above, the linearity of the input to the reconstruction on the material budget is essential for an artefact-free three-dimensional imaging. To overcome this, calibration measurements applying materials of different, known properties in terms of radiation length and thickness, can furthermore be performed and applied.

\section*{Acknowledgements}

The phantom has been obtained from \textit{Universitätsklinikum Hamburg-Eppendorf}.
This project received funding via the DESY Generator Program.
The authors acknowledge support from DESY (Hamburg, Germany), a member of the Helmholtz Association HGF, and thank all the technical groups at DESY for their work and support in the ARES implementation, maintenance and operation.

\bibliographystyle{Frontiers-Vancouver}
\bibliography{electronCT_bib}

\end{document}